\documentclass[prd,twocolumn,nofootinbib,notitlepage,aps,tightenlines,preprintnumbers,amsmath,amssymb,amsfonts,showpacs,superscriptaddress]{revtex4-1}

\RequirePackage[colorlinks=true
,urlcolor=blue
,anchorcolor=blue
,citecolor=blue
,filecolor=blue
,linkcolor=blue
,menucolor=blue
,linktocpage=true
,pdfproducer=medialab
,pdfa=true
]{hyperref}

\usepackage{amsmath,amssymb,amsthm,amsfonts}
\usepackage{graphicx,tabularx}
\usepackage{float}
\usepackage{multirow}  
\usepackage{cancel}
\usepackage{comment}

\usepackage{cleveref}
\crefname{section}{Sec.}{Secs.}
\crefname{figure}{Fig.}{Figs.}
\crefname{equation}{Eq.}{Eqs.}
\crefname{appendix}{Appendix}{Appendices}

\newcommand{\ssectionfirst}[1]{\noindent \textbf{#1}\!\!}
\newcommand{\ssection}[1]{\vspace*{0.04in}\noindent\textbf{#1}\!\!}

\newcommand{\be}{\begin{equation} \begin{aligned}}
\newcommand{\ee}{\end{aligned} \end{equation}}

\newcommand{\mev}{\text{MeV}}
\newcommand{\gev}{\text{GeV}}
\newcommand{\tev}{\text{TeV}}

\newcommand{\iab}{\text{ab}^{-1}}

\newcommand{\cm}{\text{cm}}
\newcommand{\m}{\text{m}}
\newcommand{\hz}{\text{Hz}}

\newcommand{\mrad}{\text{mrad}}

\begin{document}
\preprint{FERMILAB-PUB-20-477-AE-PPD-T}
	
\title{\mbox{FORMOSA: Looking Forward to Millicharged Dark Sectors}}

\author{Saeid Foroughi-Abari}
\email{saeidf@uvic.ca}
\affiliation{Department of Physics and Astronomy, University of Victoria, Victoria, BC V8P 5C2, Canada}

\author{Felix Kling}
\email{felixk@slac.stanford.edu}
\affiliation{SLAC National Accelerator Laboratory, 2575 Sand Hill Road, Menlo Park, CA 94025, USA}

\author{Yu-Dai Tsai}
\email{ytsai@fnal.gov}
\affiliation{Theory Department, Fermi National Accelerator Laboratory (Fermilab), Batavia, IL 60510, U.S.A.}

\begin{abstract}
We identify potentially the world's most sensitive location to search for millicharged particles in the 10 MeV to 100 GeV mass range: the forward region at the LHC. We propose constructing a scintillator-based experiment,  FORward MicrOcharge SeArch (FORMOSA) in this location, and estimate the corresponding sensitivity projection. We show that FORMOSA can discover millicharged particles in a large and unexplored parameter space, and study strongly interacting dark matter that cannot be detected by ground-based direct-detection experiments. The newly proposed LHC Forward Physics Facility (FPF) provides an ideal structure to host the full FORMOSA experiment.
\end{abstract}

\maketitle

\ssectionfirst{Introduction -}
\label{sec:intro}
Searching for millicharged particles, or MCPs, is a test for the empirical charge quantization~\cite{Dirac:1931kp} as well as predictions from motivated UV theories, including grand unified theories (GUT)~\cite{Pati:1973uk,Georgi:1974my} and string theory~\cite{Wen:1985qj,Shiu:2013wxa}). MCP can also arise as a low-energy consequence of a theory with a massless kinetic-mixing dark photon~\cite{Holdom:1985ag}. Recently, the consideration of MCP as dark matter (DM) \cite{Brahm:1989jh, Feng:2009mn, Cline:2012is} and the connection to the explanation to the EDGES anomaly \cite{Bowman:2018yin, Barkana:2018lgd, Berlin:2018bsc, Slatyer:2018aqg, Liu:2019knx} rekindle the experimental programs to look for MCPs.

Within the past few decades, the MCP probes at terrestrial experiments such as colliders, fixed-target experiments as well as astrophysical or cosmological observations have been vastly studied and searched for~\cite{Dobroliubov:1989mr, Prinz:1998ua, Davidson:2000hf, Golowich1987, Babu:1993yh, Gninenko:2006fi, Agnese:2014vxh, Haas:2014dda, Ball:2016zrp, Alvis:2018yte, Magill:2018tbb}. 

A dedicated experiment, milliQan \cite{Haas:2014dda}, was proposed at LHC that would detect MCPs produced by proton collisions using scintillator-based detectors. Later, FerMINI~\cite{Kelly:2018brz} with a similar setup was considered for proton fixed-target and neutrino experiments, primarily targeting Fermilab proton beamlines or CERN SPS beam. Recently, this idea is followed by a proposal at J-PARC proton fixed-target facilities~\cite{Choi:2020mbk}.

In general, there is a strong advantage in considering experiments in the intersection of both high-energy and high-intensity frontier, to study dark-sector or long-lived particles. Traditionally such searches are done either in the LHC transverse region or proton fixed-target experiments~\cite{Batell:2009di, Essig:2010gu, deNiverville:2011it, Kahn:2014sra, deNiverville:2015mwa, Gardner:2015wea, Izaguirre:2015pva, Pospelov:2017kep, Darme:2017glc,Magill:2018jla, Berlin:2018pwi, Magill:2018tbb, Jordan:2018gcd, deNiverville:2018dbu, Bertuzzo:2018itn, Bertuzzo:2018ftf, Ballett:2018ynz, Arguelles:2018mtc, Batell:2018fqo, Arguelles:2019xgp, Kelly:2018brz, Tsai:2019mtm, DeRomeri:2019kic, Dobrich:2019dxc, Buonocore:2019esg, Ariga:2018uku}. However, it was noted that the production rates available in the forward direction are comparable to those achieved at beam dump experiments, so one can view the detectors located at the LHC forward regions as {\it very energetic beam-dump experiments} given the high statistics one can accumulate in this region.  The LHC's forward region can be regarded as the true High-Intensity Energy frontier: this is where one can get a high flux of low-mass dark sector particles through direct productions, and meson decays. However, until recently, this region was neglected.

In this paper, we consider two scenarios studying millicharged particles in the LHC forward physics region. First, we consider installing a minimal MCP detector in (or directly move the proto-milliQan detector~\cite{Ball:2020dnx} to) the UJ12 hall, which we call FORMOSA-I for convenience). Secondly, we consider constructing a full-size milliQan-type detector in the Forward Physics Facility (FPF, an expanded UJ12 hall~\cite{SnowmassFPF}), referred to as FORMOSA-II. FORMOSA will join FASER as the second proposal to take advantage of the LHC forward physics region and be part of the FPF.

We will also demonstrate another strong advantage of FORMOSA: suppose that MCPs make up a small fraction of the total DM abundance. FORMOSA therefore provides a probe of DM that is independent of the local DM flux and insensitive to the attenuation for the MCP passing through materials, given the energy of the DM particles produced in LHC. We will show that FORMOSA can cover most of the ``millicharged strongly interacting dark matter (SIDM)" window, as discussed in~\cite{Emken:2019tni, Plestid:2020kdm}. FORMOSA can also study other beyond standard model (BSM) scenarios, such as heavy neutrinos and DM with sizeable electric dipoles \cite{Sher:2017wya,Chu:2018qrm,Frank:2019pgk,Chu:2020ysb}.

\ssection{Location -}
\label{sec:location}
We propose FORMOSA to be located in the far-forward direction, close to the beam collision axis, where it can benefit from an enhanced MCP production cross-section compared to the transverse direction. 

\begin{figure*}[t]
\centering
\includegraphics[clip, trim=20mm 10mm 0mm 15mm, width=0.48\textwidth]{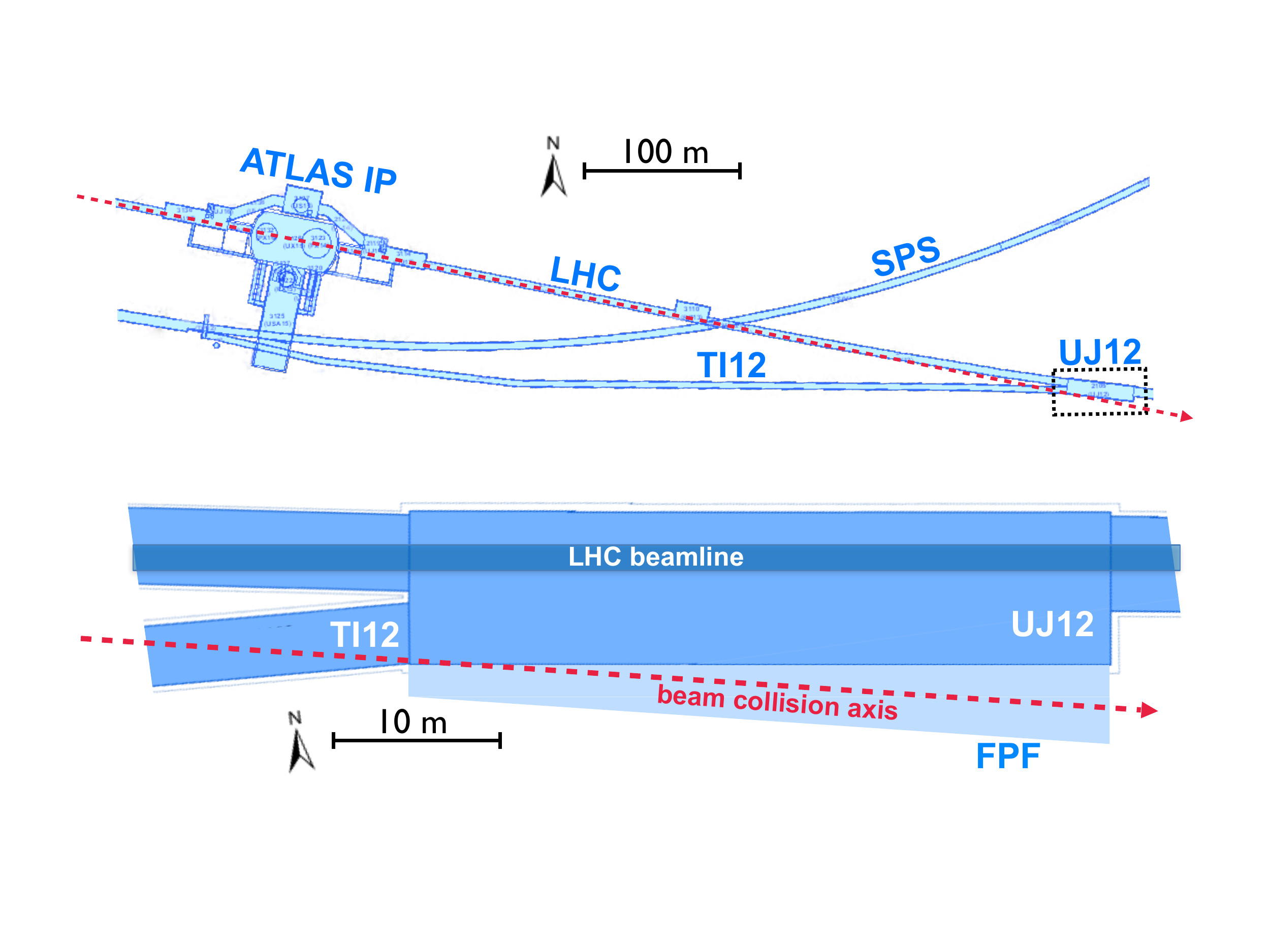} 
\includegraphics[width=0.48\textwidth]{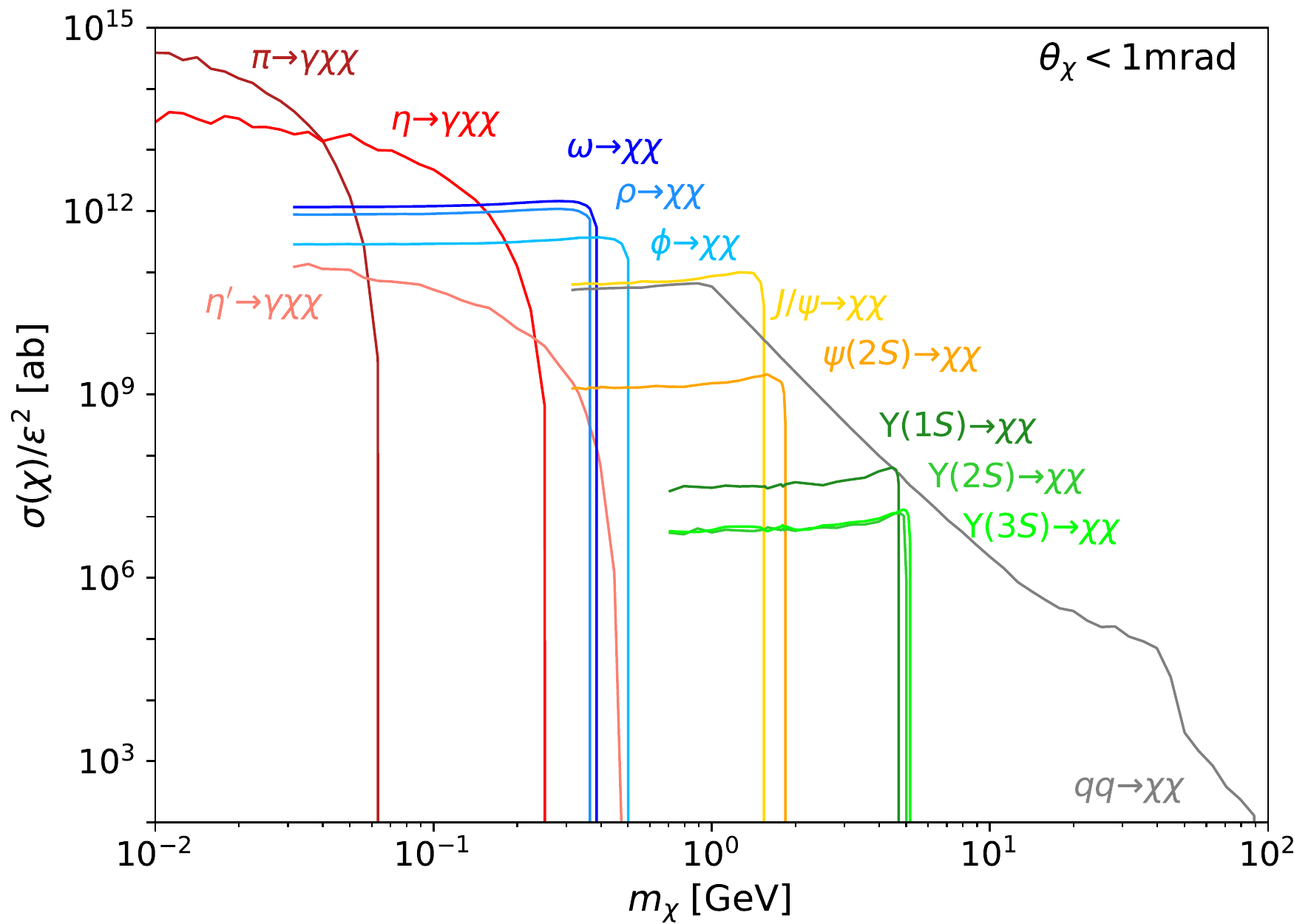} 
\caption{\textbf{Left:} The proposed location of FORMOSA in the cavern UJ12 or side tunnel TI12 (blue) close to the beam collision axis (red). The Forward Physics Facility (FPF) extension is shown as a light-blue area. 
\textbf{Right:} Production cross section of MCP in the forward direction, $\theta_\chi < 1~\mrad$, for different production modes and MCP masses. \vspace{-0.5cm}}
\label{fig:production}
\end{figure*}

A suitable location is available roughly $500~\m$ downstream from the ATLAS interaction point in the cavern UJ12 or the tunnel TI12, shown in the left panel of \cref{fig:production}. One can also consider the nearly symmetrical TI18 and UJ18 on the opposite side of ATLAS. During Run~3 of the LHC, TI12 will host the FASER experiment to search for long-lived particles~\cite{Feng:2017uoz, Feng:2017vli, Kling:2018wct, Feng:2018pew, Berlin:2018jbm, Ariga:2018pin, Ariga:2018uku, Ariga:2018zuc, Kling:2020mch}, and study neutrino interactions~\cite{Abreu:2019yak, Abreu:2020ddv}, with continuations being proposed for the HL-LHC era~\cite{SnowmassFASER2, SnowmassFASERnu2}. For this reason, TI12 and UJ12 are equipped with lighting, power, stairs, and support structures to safely transport detector components around the LHC. Recently, it has also been proposed to enlarge the UJ12 cavern to create a FPF, which could house FORMOSA and other forward experiments~\cite{SnowmassFPF}. 

TI12 and UJ12 are shielded from the ATLAS IP by the forward LHC infrastructure, consisting of magnets and absorbers, as well as $100~\m$ of rock. Particle fluxes and radiation levels have been simulated using FLUKA~\cite{FLUKAstudy} and validated experimentally~\cite{Ariga:2018pin, Abreu:2020ddv}, indicating that the particle fluxes at this location are sufficiently low. 

\ssection{MCP Production -} 
\label{sec:production}
We study MCP $\chi$ with electric
charge $Q_\chi$, mass $m_\chi$ and define $\epsilon \equiv Q_\chi/e$. The small electric charge can come from directly introducing a tiny $U(1)$ hypercharge to $\chi$. It can also be generated by considering a massless dark photon, kinetically mixed with SM, that couples to $\chi$ and induce the millicharge of $\chi$ in a convenient basis \cite{Okun:1982xi,Holdom:1985ag}. 

We perform a dedicated Monte Carlo study to estimate the flux of MCP produced at the LHC. The different production channels and their corresponding production cross sections in the forward direction, within $\theta_\chi<1~\mrad$ of the beam collision axis, are summarized in the right panel of \cref{fig:production}. 

If the MCP is light, it is primarily produced in both pseudoscalar meson decays such as $\pi^0 \to \gamma\chi\chi $ and vector meson decays such as $\omega, J/\psi,\Upsilon \to \chi\chi$. We generate the spectra of light mesons using \texttt{EPOS-LHC}~\cite{Pierog:2013ria}. The spectra of the Charmonium and Bottomonium states are simulated using \texttt{Pythia~8}~\cite{Sjostrand:2014zea}, which we have turned to their spectra measured at LHCb~\cite{Aaij:2015rla, Aaij:2018pfp, Aaij:2019wfo}. More details on the simulation and validation of forward meson production at the LHC and their decay into MCPs can be found in the supplementary material. 

Heavy MCP's are primarily produced in partonic scattering $qq \to \chi\chi$. We simulate this Drell-Yan production mode using \texttt{MadGraph~5}~\cite{Alwall:2014hca} and \texttt{Pythia~8}. To ensure that the PDFs are well defined, we require the invariant mass of the MCP pair to be larger than $2~\gev$, which leads to a constant MCP production cross sections for $m_\chi < 1~\gev$.

Before proceeding, let us note that particle production rates are enhanced in the forward direction. Feynman scaling arguments tell us that MCP production is approximately flat is rapidity, $d N / dy \approx \rm{const.}$~\cite{GrosseOetringhaus:2009kz}. When $m_{\chi}\ll p_{T}$, we can further approximate the rapidity $y$ with the pseudorapidity $\eta = -\log(\tan(\theta/2))$ and obtain a purely geometric statement $dN / d\eta \approx \rm{const.}$. Let us now compare the flux of MCPs going through a $1~\m \times 1~\m$ area at i) a transverse location (T) similar to milliQan at $\eta \sim 0$ about $30~\m$ from the beam axis, and ii) a forward location (F) about $500~\m$ downstream and off-set by $2~\m$ from the beam axis, corresponding to $\eta \sim 6$. We find $N_{\rm F}/N_{\rm T}\sim \Delta \eta_{F}/\Delta \eta_{T}\times\Delta \phi_{\rm F}/\Delta \phi_{\rm T}\sim 250$, meaning that the forward particle flux is indeed strongly enhanced.

\ssection{Detector and Signature -} 
\label{sec:detector}
We will consider a minimal detector (FORMOSA-I) and a full MCP detector (FORMOSA-II). For FORMOSA-I we consider a setup similar to the milliQan demonstrator with size $0.2~\m \times 0.2~\m \times 4~\m$, which consists of 4 layers each containing 16 scintillator bars coupled to a PMT. FORMOSA-II would be a $1~\m \times 1~\m \times 4~\m$ array consisting of 4 layers of 400 scintillator bars. We consider FORMOSA-I to be located in the UJ12/TI12 hall and 2 meters off-axis, and we consider FORMOSA-II to be on-axis and located in the proposed FPF (expanded UJ12 hall).

When a MCP particle traverses the detector, photoelectrons (PE) are produced due to the ionization energy deposition directly within each stack of scintillator. The mean rate of energy deposition $\langle -dE/dx\rangle$ described by Bethe-Bloch equation scales as $\epsilon^2$ with a typical value of $\sim 2 \, \epsilon^2 \, \mev/\cm$ in a material of density 1 g/$\mathrm{cm}^3$ for a minimum-ionizing charged particle~\cite{Tanabashi:2018oca}. The average number of PEs collected with the detection efficiency $\varepsilon_{\rm det}$ in a scintillator bar with the density $\rho_{s}$ and the length $L_{s}$ can be estimated as $\bar{N}_{\rm PE}\approx \varepsilon_{\rm det}\rho_{s}L_{s}{\times}\langle -dE/dx\rangle{\times}Y_{\gamma}$, where $Y_{\gamma}\sim 1.1\times 10^4~\mev^{-1}$ is the photon yield deposited in the typical plastic scintillators~\cite{SaintGobain:2016}. In performing calculations of  $\bar{N}_{\rm PE}$, we use the mean rate of energy loss formula, given in PDG~\cite{Tanabashi:2018oca}, as a function of the MCP's energy. We assume a detection efficiency $\epsilon \approx 10\%$, as considered by the milliQan collaboration~\cite{Ball:2016zrp}. 

For FORMOSA, we will search for a quadruple coincidence of hits with $\bar{N}_{\rm PE}\ge 1$ within a 20 ns time window. The detection of a millicharged particle requires at least one PE in each stack of the scintillator. Therefore, the probability of observing a MCP in a detector consisting of 4 layers follows the Poisson distribution, $P_{\rm det}=(1-\exp(-\bar{N}_{\rm PE}))^4$. Considering the number of MCPs passing through the detector $N_\chi$, the total number of signal events is $N_\chi \cdot P_{\rm det}$.

\ssection{Background -} 
\label{sec:backgrounds} 
The potential background sources of an MCP detector at the forward physics region can be classified as beam related and beam unrelated. According to the proto-milliQan study~\cite{Ball:2020dnx}, in the transverse region, the beam-unrelated background dominates, and it can be controlled by adding an additional layer to the original three-layer millQan design. However, in the forward region at the LHC considered in this paper, we found that the beam-related background could be much more important. The background reduction strategy and the choices of PMT's need to be reconsidered accordingly. We will first review the beam unrelated background and its reduction strategy, and then discuss the beam-related background.

The beam unrelated background of FORMOSA can come from cosmic muons and dark current, and their combined coincident signatures. The cosmic muons interacting with the cavern walls generate a spray of electrons and gamma rays. The produced shower causing a scintillation in the detector is a significant background source for the proto-milliQan detector~\cite{Ball:2020dnx}. In addition, due to the random emission of thermal electrons from the photo-cathode, dark current pulses can be produced in each PMT. These two sources of detector background lead to the similar signature to MCPs.

Several techniques are developed to reduce these beam-unrelated background events, ~\cite{Haas:2014dda,Ball:2016zrp}. These strategies can be summarized as requiring multiple-coincidence as a detection signature, implementing a veto of large-PE events, shifting or enlarging the middle detector array (to eliminate coincident low-PE pulses caused by energetic muons scraping through the surface of the scintillator layers), and considering a dead-time veto of the afterpulses. In addition to scintillator bars, the milliQan prototype design has considered components such as the scintillator panels and four scintillator slabs along the length of the detector for further reduction of backgrounds~\cite{Ball:2020dnx}.  The scintillator panels are used as shields for the bars from the top and sides to reject backgrounds due to cosmic muon showers. The slabs provide time information, shielding from neutron radiation, and help to veto deposits due to beam and cosmic muons passing the bars. As demonstrated by the milliQan collaboration, the beam unrelated background can be reduced to nearly zero with all these background reduction strategies. 

Here, we show an analytical estimation of the dark-current background rate to demonstrate the strength of the quadruple coincidence requirement. The rate of random coincidence of 4 such dark pulses takes the form of $R = B^4 \tau^{3}\simeq 5\times 10^{-13}$ Hz, assuming a typical dark current rate of $B=500$ Hz~\cite{Haas:2014dda} and a quadruple-incidence time window of $\tau=20$ ns. 
Accounting for the number of bars in each layer $N$, the total dark-current background rate of the detector is $N\times R$. Assuming an instantaneous luminosity of $\sim 10^{35}$ cm$^{-2}$s$^{-1}$, the HL-LHC will deliver $3~\iab$ in an approximate trigger live-time of one year. The FORMOSA-II full detector consists of 400 PMTs per stack would have a total dark-current background rate of $\sim 0.01$ events per year. This rate is greatly suppressed compared to the original milliQan detector with triple coincidence, which is expected to have $\sim 300$ background events per year~\cite{Ball:2016zrp, Kelly:2018brz}. The same conclusion taking into account other beam unrelated background is also shown in \cite{MilliQan:2020update}. 

A new challenge arises for the dedicated MCP search in for forward physics region, given a large flux of high-energy muons from the beam collisions. For FORMOSA-I, we consider a location inside the cavern UJ12, where a muon rate $\lesssim 1~\hz/\cm^{2}$ can be achieved~\cite{Abreu:2019yak} (one can consider the same muon rate for FORMOSA-II). These energetic muons (and the secondary particles they produce inside the detector) can cause large pulses in the PMT. We will implement online-vetos of large-PE pulses to avoid the readout deadtime and ensure a high signal efficiency. 

In addition, afterpulses, which can appear with a delay time of a few $\mu$s after the initial pulse, could become sources of background events. These smaller pulses can appear correlated in the $\tau$ time window. Thus, they may be indistinguishable from the small PE events and cannot be vetoed by the large PE cuts. However, the rate of the afterpulses with a delay time of $\delta t \gtrsim 10~\mu$s drops below the dark current rate for most of the PMTs~\cite{Aiello:2018nvl, TheDEAP:2017bxf}. We can therefore remove the afterpulse background would using a veto: assuming one muon every 100 $\mu$s for FORMOSA-II and an afterpulsing duration of $\sim 10~\mu$s, roughly $10\%$ of the data needs to be vetoed, resulting in a live-time efficiency of $\sim 0.9$. For FORMOSA-I, this is a much smaller issue since it has a smaller detector area and thus a lower total muon rate. Considering better PMTs with reduced afterpulse duration can improve the live-time efficiencies for both FORMOSA-I and II.

Other beam-related background sources come from beam gas collisions and the beam halo collisions with the beam pipe. Some of these radiation backgrounds have been measured~\cite{Ariga:2018pin} and can be used to estimate the radiation level. However, requiring of the reconstructed pulses to be pointing to the ATLAS IP, and the use of the panel veto-shields, should reduce this type of beam-related background to a negligible level. 

\begin{figure*}[t]
\centering
\includegraphics[width=0.49\textwidth]{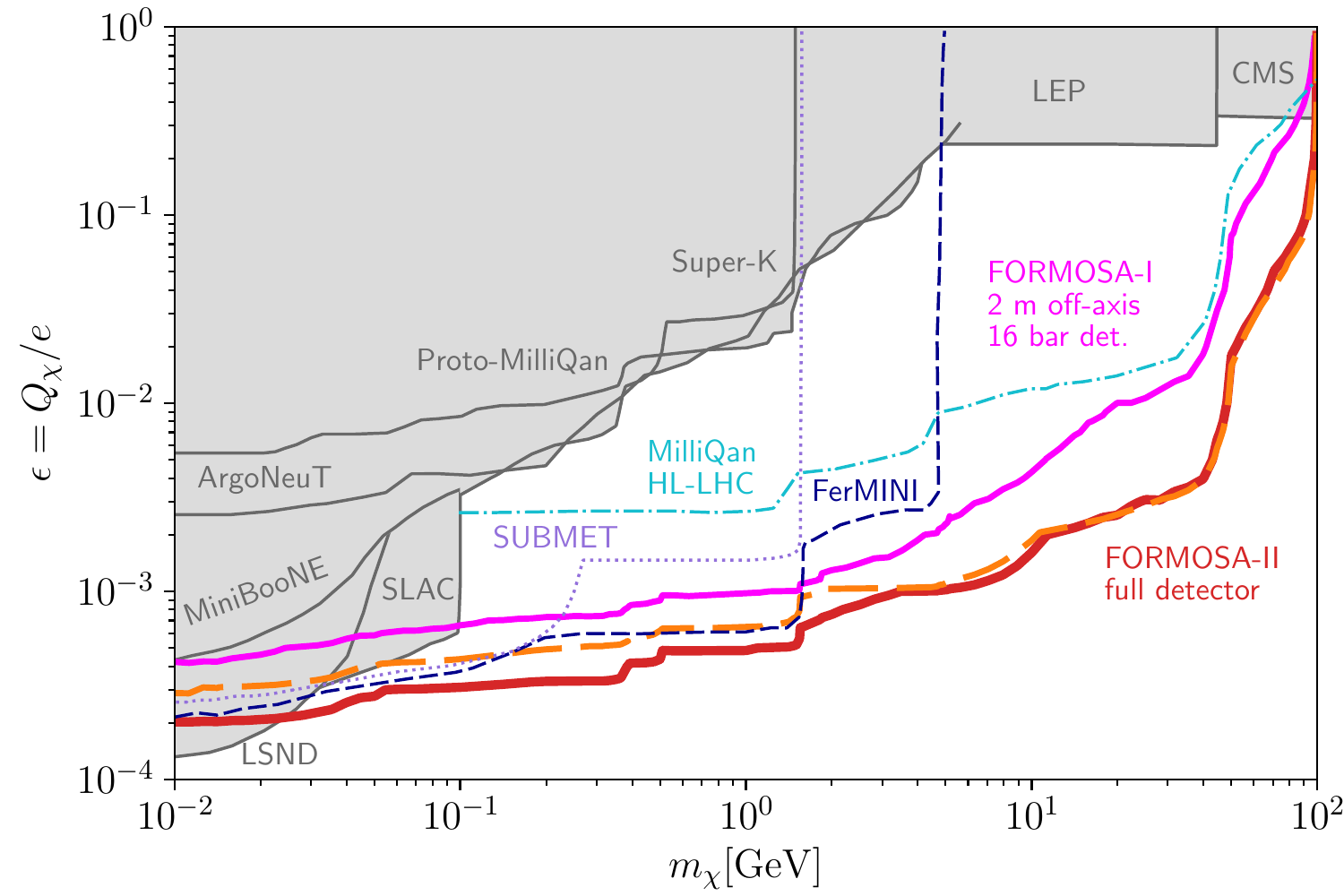}
\includegraphics[width=0.49\textwidth]{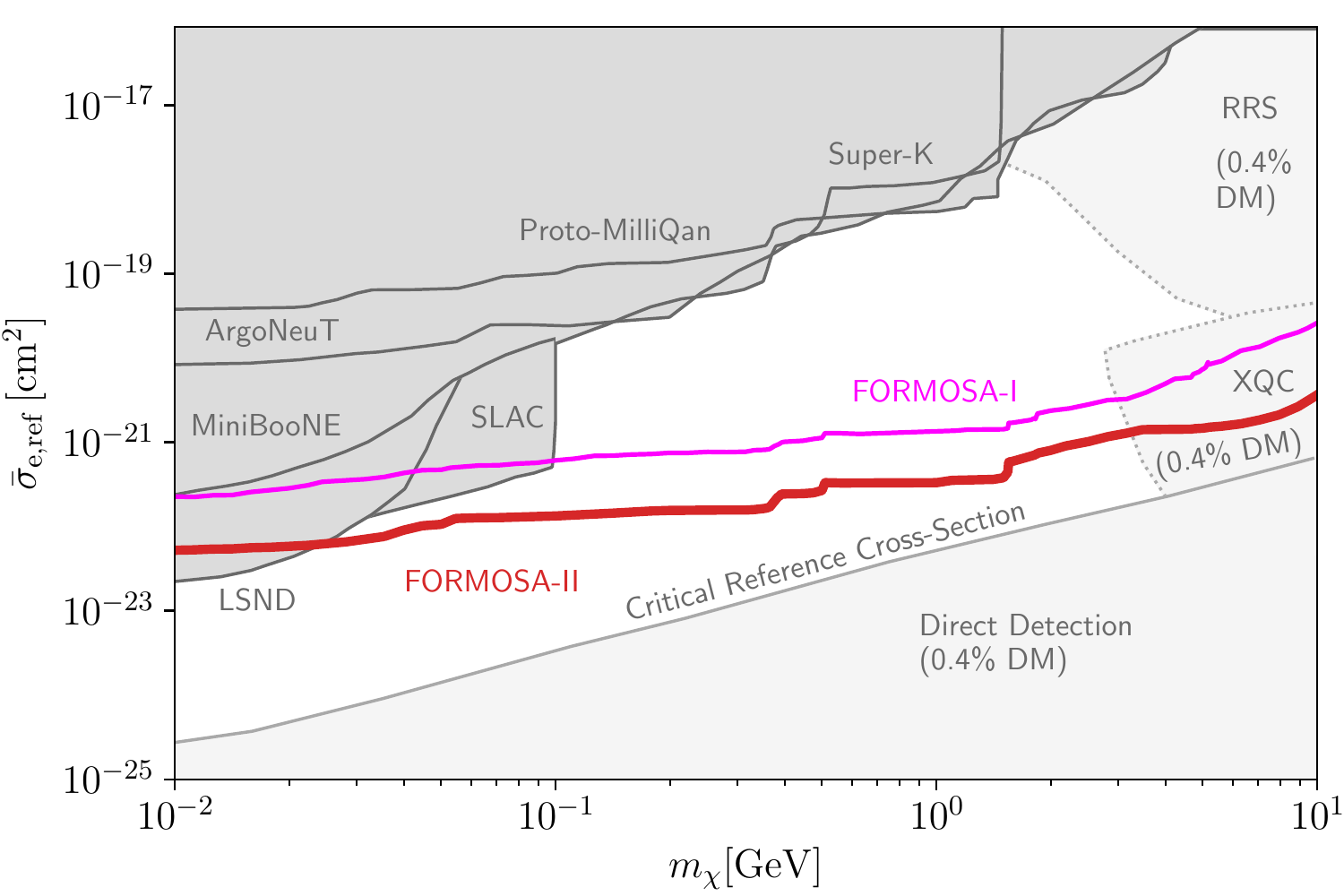}
\caption{\textbf{Left:} Parameter space for MCPs shown in the plane of the fractional charge $\epsilon$ versus MCP mass $m_{\chi}$. The sensitivity reach of FORMOSA-I shown as a solid magenta curve corresponds to placing the minimal detector (16 bars per layer) 2 m off-axis at the LHC UJ12 hall, assuming 3000 fb$^{-1}$ of integrated luminosity. The solid (dashed) red line shows the expected sensitivity of FORMOSA-II on-axis (2 m off-axis) at FPF (expanded UJ12 hall), assuming the same luminosity. Exclusions from other previous studies are shown in gray and 
projections for milliQan HL-LHC~\cite{Ball:2016zrp} (dashed cyan), FerMINI at LBNF/DUNE~\cite{Kelly:2018brz} (dashed blue) and SUBMET at J-PARC (dashed purple)~\cite{Choi:2020mbk} are shown for comparison (see the text for further details). 
\textbf{Right:} The sensitivity reaches of FORMOSA-II in the millicharged SIDM window is shown in terms of the reference cross-section $\bar{\sigma}_{\rm e,ref}$ defined in the text. In addition to accelerator constraints we show constraints from direct-detection experiments (assuming $0.4~\%$ DM abundance for the direct-detection experiments)~\cite{Emken:2019tni}. 
} 
\label{fig:sensitivity}
\vspace{-0.5cm}
\end{figure*}
 
\ssection{Sensitivity -} 
\label{sec:Sensitivity}
In the left panel of \cref{fig:sensitivity}, we present the projected sensitivity reaches of FORMOSA-I and II in terms of the MCP mass $m_\chi$ and charge ratio $\epsilon\equiv Q_{\chi}/e$, assuming an integrated luminosity of $3~\iab$ at the HL-LHC. We require 3 signal events, and a 0.9 live-time efficiency. The choice of 3 events is based on the discussions in the background section that the beam-unrelated background can be reduced to a negligible level with the requirement of quadruple coincidence. We also assume the beam-produced muon background can be reduced to a negligible level if one install suitable PMTs and apply appropriate cuts to eliminate the afterpulse backgrounds. These sensitivity projections can be easily adjusted with more realistic background determinations. The line corresponds to FORMOSA-I placed $2~\m$ off-axis is colored magenta. FORMOSA-II on-axis ($2~\m$ off-axis) is plotted red (orange and dashed). 

We plot the existing constraints as gray-shaded regions. These include bounds from SLAC~\cite{Prinz:1998ua}, LEP~\cite{Akers:1995az, Davidson:2000hf}, CMS~\cite{CMS:2012xi, Jaeckel:2012yz}, LSND and MiniBooNE~\cite{Magill:2018tbb}, ArgoNeuT at Fermilab~\cite{Acciarri:2019jly}, proto-milliQan at LHC~\cite{Ball:2020dnx}, and cosmic-ray produced MCP constraint from Super-K~\cite{Plestid:2020kdm}. For comparison, we also show the sensitivity projections from the full milliQan experiment at $14~\tev$ LCH with $3~\iab$ integrated luminsoity \cite{Ball:2016zrp}, the proposed FerMINI experiment at LBNF/DUNE with a beam energy of $120~\gev$ and $10^{21}$ POT~\cite{Kelly:2018brz}, and the proposed SUBMET experiment assuming $10^{22}$ POT at the $30~\gev$ proton beam at J-PARC~\cite{Choi:2020mbk}. 

Base on our analysis, one can see that FORMOSA-I, a low-cost minimal detector to be placed in UJ12, would provide a better sensitivity reach in comparison to the full MilliQan run. As discussed above, this is due to an enhanced flux of MCP in the forward region, in comparison to that of the transverse region. Up to $\mathcal{O}(10^6)$ signal events at $m_{\chi}=1~\gev$ near the proto-milliQan bound are possible.

Also note that, although we derive our sensitivity projections based on the luminosity of HL-LHC, it could be possible to employ a similar setup already during Run 3 of the LHC. In particular, one could install the proto-milliQan detector in the TI12 cavern and perform a  test run for FORMOSA to better understand the detector environment and experimental challenges.

We further show that FORMOSA-II provides leading sensitivity projections in a large window of MCP mass from $100~\mev$ to $100~\gev$, exceeding the reaches of other similar proposals. We also see that, even if the detector is placed a few meters off the beam axis, the sensitivity would not be strongly affected. This allows to directly place FORMOSA inside the existing cavern UJ12. 

The features of the sensitivity projections of FORMOSA can be easily understood. In the high mass region, the detection probability is $(1-e^{-N_{\rm PE}})^4 \sim 1$, and the number of produced MCPs controls the signal event rate. In contrast, in the low mass region and for $\epsilon \lesssim 10^{-3}$, we expect a small number of PEs, leading to a sharp drop in the number of signal events as we reach the detection limit of the scintillator \cite{Kelly:2018brz,Choi:2020mbk}.

\ssection{Millicharged Strongly-Interacting DM -} 
\label{sec:sidm}
MCP in our parameter of interest can account for a fraction of the dark matter (DM) abundance, and cannot be detected by the direct-detection experiments when the cross-section is larger than certain critical values (derived in~\cite{Emken:2019tni}). The ambient DM with a substantial cross-section with Standard Model (SM) particles can lose most of its kinetic energy through the interactions with SM particles. For some model parameters, the DM particles lose most of their energy and hence cannot be detected by ground-based direct detection experiments after interacting with the atmospheric particles and the crust. These DM particles are generally referred to as strongly interacting DM (SIDM)~\cite{Rich:1987st, Starkman:1990nj, Starkman:212913}. In~\cite{Emken:2019tni, Plestid:2020kdm}, an unconstrained region of parameter space is identified, which can be referred to as a millicharged SIDM window, and FORMOSA can provide strong sensitivity in this parameter space. 

In the right panel of \cref{fig:sensitivity}, we show our results in terms of $m_\chi$ and the conventional ``reference cross section'' $\bar{\sigma}_{\rm e,ref} = 16 \pi \alpha^2 \epsilon^2 \mu^2_{\chi e}/q^4_{ \rm d, ref}$. Here, $q_{\rm d, ref}$ is chosen to be the typical momentum transfer in $\chi-e$ scatterings for semiconductor or noble-liquid targets (taken to be $\alpha m_e$~\cite{Emken:2019tni}) and $\mu_{\chi e}$ is the reduced mass of the electron and $\chi$. We do not plot the constraints based on the millicharged DM accelerated by astrophysical sources~\cite{Hu:2016xas, Dunsky:2018mqs, Li:2020wyl, Chuzhoy:2008zy}, as they require extra assumptions beyond local DM properties. Note that we assume 0.4 $\%$ of the DM to be MCP, for all the direct-detection constraints, to avoid strong cosmological constraints \cite{Dubovsky:2003yn, Dolgov:2013una, Kovetz:2018zan}.

We show that FORMOSA can help cover a large part of the millicharged DM region that is previously unconstrained. Our study also demonstrates two strong advantages of accelerator probes of DM in general: (i) the accelerator probes are not sensitive to the material's attenuation, given that the particles produced from beam interactions have large kinetic energy (unlike the ambient DM, which has much lower kinetic energy), and (ii) the accelerator constraints are independent of the fractional composition of DM (unlike the direct-detection or cosmological probes~\cite{Dubovsky:2003yn, Dolgov:2013una, Kovetz:2018zan}).

\ssection{Conclusion -} 
\label{sec:conclusion}
In this letter, we consider one of the world's most sensitive setups for MCP searches, located in the forward region of the LHC interaction point. FORMOSA, a milliQan-like experiment placed $\sim 500$ downstream from ATLAS, would take advantage of enhanced MCP production in the forward direction and can provide leading sensitivity to MCPs in the 10 MeV to 100 GeV mass window. 

We also found that, unlike for the current milliQan location, beam-related backgrounds associated with the sizable flux of forward muons, such as PMT afterpulses, become important in the forward direction. This motivates additional detector design considerations, such as the use of PMTs with low afterpulse duration or the application of vetoes to control these afterpulse background. 

In addition to MCPs, FORMOSA can provide great sensitivity for other beyond standard model (BSM) particles, such as heavy neutrinos and DM with electric dipoles \cite{Sher:2017wya,Chu:2018qrm,Frank:2019pgk,Chu:2020ysb}. Furthermore, FORMOSA's location in the far forward direction allows further extensions of its physics objectives. For example, one would expect about $10^6$ neutrino interactions with the FORMOSA-II detector, providing additional opportunities for neutrino physic~\cite{ Abreu:2019yak, Bai:2020ukz, SnowmassFASERnu2} and forward particle production measurements, and indicating the physics potential of FORMOSA and the Forward Physics Facility that remains to be explored.

\newpage
\acknowledgments
We would like to thank Matthew Citron, Patrick de Niverville, Jonathan Feng, and Christopher Hill for useful discussions. We also thank Matthew Citron, Jonathan Feng and Adam Ritz for valuable and detailed comments on the draft. We are grateful to the authors and maintainers of many open-source software packages, including
\texttt{CRMC} \cite{CRMC},
\texttt{EPOS-LHC} \cite{Pierog:2013ria},
\texttt{Jupyter} notebooks \cite{soton403913}, 
\texttt{MadGraph~5} \cite{Alwall:2014hca},
\texttt{Matplotlib} \cite{Hunter:2007}, 
\texttt{NumPy} \cite{numpy}, 
\texttt{PyHepMC} \cite{buckley_andy_2018_1304136},
\texttt{Pythia~8} \cite{Sjostrand:2014zea} and
\texttt{scikit-hep} \cite{Rodrigues:2019nct}. 

F.K. is supported by U.\,S.~Department of Energy grant DE-AC02-76SF00515. The work of S.F. was supported in part by NSERC, Canada. Part of this document was prepared by Y.-D.T. using the resources of the Fermi National Accelerator Laboratory (Fermilab), a U.S. Department of Energy, Office of Science, HEP User Facility. Fermilab is managed by Fermi Research Alliance, LLC (FRA), acting under Contract No. DE-AC02-07CH11359. 

\appendix
\vspace{5mm}
{\centering \bf Supplementary Material\\ }
\vspace{5mm}
\ssectionfirst{MCP Production in Meson Decays -}
\label{app:mesons}
If the MCP $\chi$ is sufficiently light, $m_\chi \lesssim 5~\gev$, it can be produced in the decay of SM mesons $M$.  This requires a reliable description of forward particle production, best validated with or tuned to available data. 

\begin{figure*}[t]
\centering
\includegraphics[width=0.32\textwidth]{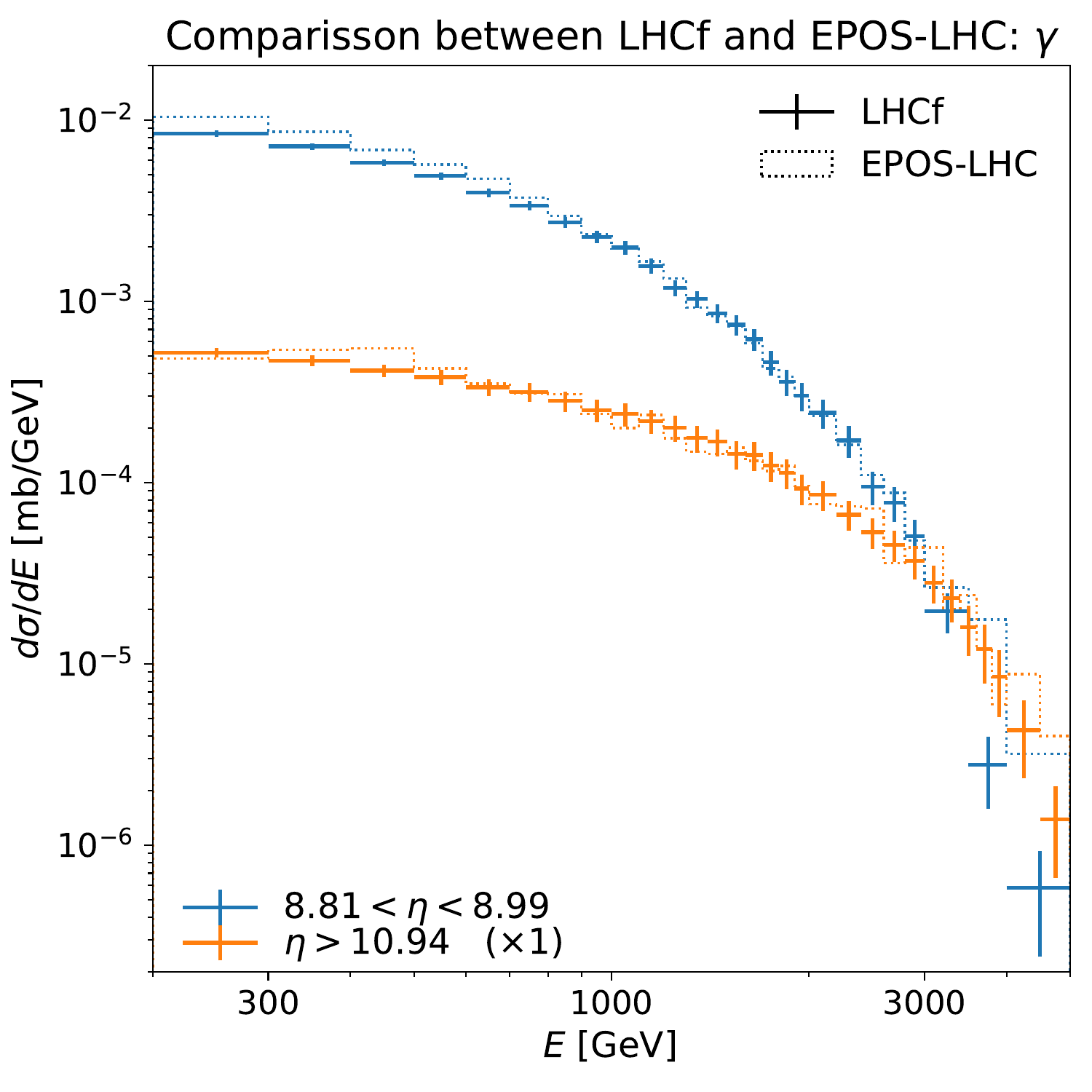}  
\includegraphics[width=0.32\textwidth]{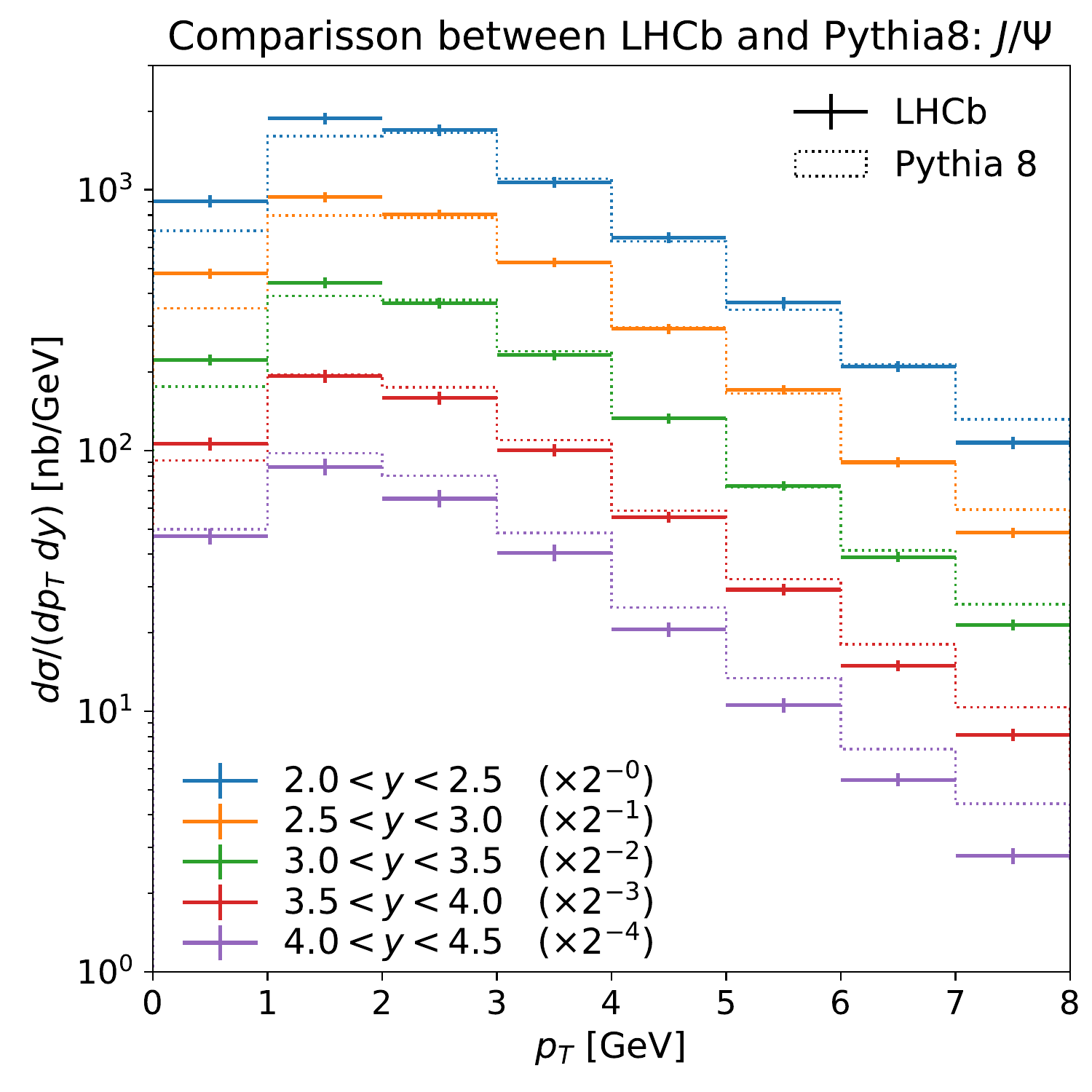}  
\includegraphics[width=0.32\textwidth]{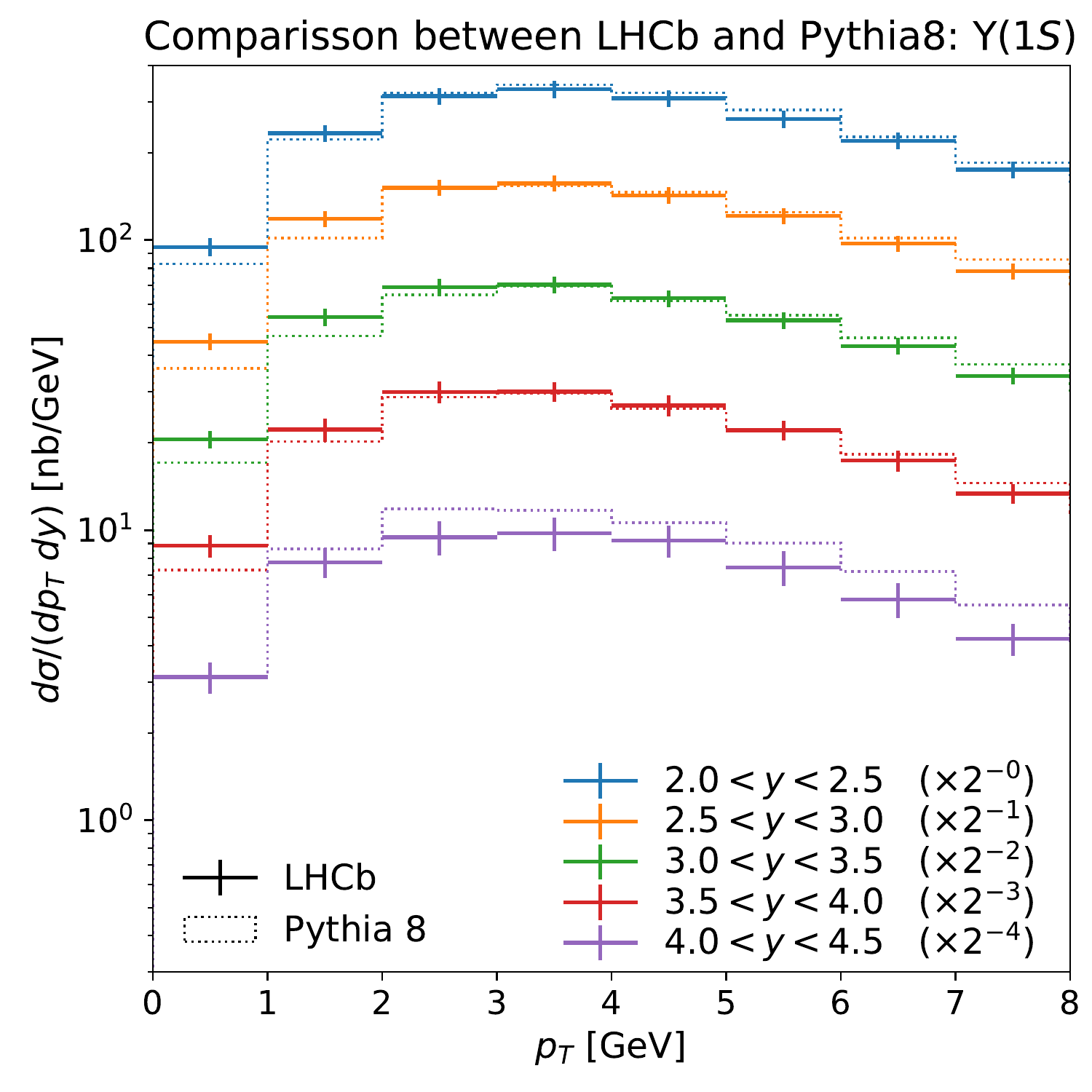} 
\caption{Validation of forward meson production. Forward photon production predicted by \texttt{EPOS-LHC} (left), $J/\psi$ production (center) and $\Upsilon(1S)$ production predicted by \texttt{Pythia~8} compared to measurements by LHCf~\cite{Adriani:2017jys} and LHCb~\cite{Aaij:2015rla, Aaij:2018pfp} at $13~\tev$ LHC.}
\label{fig:validation}
\end{figure*}

The production of light mesons, $M=\pi^0, \eta, \eta', \omega, \rho$ and $\phi$, is simulated using \texttt{EPOS-LHC}~\cite{Pierog:2013ria}, as implemented in the simulation package \texttt{CRMC}~\cite{CRMC}, which is a dedicated Monte-Carlo generator designed to describe minimum bias hadronic interactions at both particle colliders and cosmic ray experiments.  In the left panel of \cref{fig:validation} we compare the predicted energy spectra of far forward photons (mainly from $\pi^0$ decay) produced in $13~\tev$ collisions to those measured at LHCf~\cite{Adriani:2017jys}, and find good agreement over the full spectrum. 

In addition, we simulate the production of Charmonium and Bottomonium using \texttt{Pythia~8}~\cite{Sjostrand:2014zea}. Since the default setup of \texttt{Pythia~8} tends to overestimate their production rate at small transverse momenta $p_T$, we use the predefined \texttt{SuppressSmallPT} user hook to suppresses the production rate by a factor 
\be
 \left( \frac{p_T^2}{k^2 p_{T0}^2 + p_T^2 } \right)^2  \left( \frac{\alpha_S(k^2 p_{T0}^2 + Q^2_\text{ren})}{\alpha_S( Q^2_\text{ren}) } \right)^n \ ,
\ee
where $p_{T0}$ is the same energy-dependent dampening scale as used for multiparton interactions and $Q_{ren}$ is the renormalization scale. Good agreement between the simulation and measurements at LHCb~\cite{Aaij:2015rla, Aaij:2018pfp, Aaij:2019wfo} are obtained for $k=0.35$ and $n=3$, as shown in the center and right panel of \cref{fig:validation}.

In the next step, we use a MC simulation to subsequently decay the mesons into MCPs $\chi$. The pseudoscalar mesons $M=\pi^0, \eta, \eta'$ can undergo the 3-body decay $M \to  \gamma\chi \bar\chi$ . Following Ref.~\cite{DeRomeri:2019kic, Jodlowski:2019ycu}, the differential branching fraction for this process is given by 
\be
&\frac{d\text{BR}(M \!\to\! \gamma\chi\chi)}{ds \, d\!\cos\theta}
 = \frac{\epsilon^2 \alpha}{4 \pi s} \Big[1\!-\!\frac{s}{m_M^2}\Big]^3 \Big[1-\frac{4m_\chi^2}{s}\Big]^{\frac12} \\
 & \quad \times \Big[2 -  \Big(1-\frac{4m_\chi^2}{s} \Big) \sin^2 \theta \Big]  \times \text{BR}(M \!\to \!\gamma\gamma)
\ee
where $s = (p_\chi + p_{\bar\chi})^2$ is the invariant mass of the off-shell photon producing the MCP pair and $\theta$ is the angle between the momentum of $\chi$ in the off-shell photon's rest frame and the boost direction of the off-shell photon.

In addition, the vector mesons $M = \rho, \omega, \phi, J/\psi, \psi(2S)$ and $\Upsilon(nS)$ can decay directly into a pair of MCPs, $M \to \chi \bar\chi$. Following Ref.~\cite{Kelly:2018brz}, the corresponding branching fraction is given by
\be
\!\!\frac{\text{BR}(M \!\to\! \chi\bar\chi)}{\text{BR}(M \!\to\! ee) } 
= \epsilon^2 
\frac{(m_M^2 \!+\!2m_\chi^2)(m_M^2 \!-\! 4m_\chi^2)^{1/2}} {(m_M^2 \!+\! 2m_e^2)(m_M^2 \!-\! 4m_e^2)^{1/2}} \ . 
\ee

\bibliography{references}
\end{document}